\documentclass[10pt]{article}
\title{Compositional Semantics \\
       for the Procedural Interpretation of Logic
}

\date{M.H. van Emden}
\author{Research Report DCS-307-IR\footnote{
        Shortened version to appear in the Proceedings of the
        2006 International Conference on Logic Programming.}\\
        Department of Computer Science\\
        University of Victoria
          }

\begin{document}
\maketitle

\abstract{
Semantics of logic programs has been given by proof theory, model
theory and by fixpoint of the immediate-consequence operator.  If
clausal logic is a programming language, then it should also have
a compositional semantics.  Compositional semantics for programming
languages follows the abstract syntax of programs, composing the
meaning of a unit by a mathematical operation on the meanings of
its constituent units.  The procedural interpretation of logic has
only yielded an incomplete abstract syntax for logic programs. We
complete it and use the result as basis of a compositional semantics.
We present for comparison Tarski's algebraization of first-order
predicate logic, which is in substance the compositional semantics
for his choice of syntax.  We characterize our semantics by equivalence
with the immediate-consequence operator.

}
\newcommand{\cpp}{\hbox{{\tt C++}}}

\newtheorem{theorem}{Theorem}{}
\newtheorem{definition}{Definition}{}
\newtheorem{example}{Example}{}
\newtheorem{lemma}{Lemma}{}

\newcommand{\Emin}{\ensuremath{E_{\mbox{min}}}} 
\newcommand{\Emax}{\ensuremath{E_{\mbox{max}}}} 

\newcommand{\Nat}{\ensuremath{\mbox{\textbf{N}}}} 
\newcommand{\Int}{\ensuremath{\mbox{\textbf{Z}}}} 
\newcommand{\Rat}{\ensuremath{\mbox{\textbf{Q}}}} 
\newcommand{\Rea}{\ensuremath{\mbox{\textbf{R}}}} 
\newcommand{\ExtRe}{\ensuremath{\mbox{\textbf{R}}^{++}}} 
\newcommand{\Flpt}{\ensuremath{\mbox{\textbf{F}}}} 

\newcommand{\Alp}{\ensuremath{\mbox{\textbf{A}}}} 
\newcommand{\AlpStar} {\ensuremath{\Alp ^\ast}}

\newcommand{\A}{\ensuremath{\mathcal{A}}} 
\newcommand{\Herb}{\ensuremath{\mathcal{H}}} 
\newcommand{\M}{\ensuremath{\mathcal{M}}} 
\newcommand{\MI}{\ensuremath{\mathcal{M}_I}} 
\newcommand{\Pred}{\ensuremath{\mathcal{P}}} 
\newcommand{\T}{\ensuremath{\mathcal{T}}} 
\newcommand{\Var}{\ensuremath{\mathcal{V}}} 
\newcommand{\cart}{\ensuremath{\mbox{\textsc{cart}}}} 
\newcommand{\apl}{\ensuremath{\mbox{\textsc{apl}}}} 

\newcommand{\va}{{$a$}}
\newcommand{\vb}{{$b$}}
\newcommand{\vfa}{{$f(a)$}}
\newcommand{\vfb}{{$f(b)$}}

\newcommand{\pair}[2]{\ensuremath{\langle #1,#2 \rangle}}
\newcommand{\triple}[3]{\ensuremath{\langle #1,#2,#3 \rangle}}
\newcommand{\vc}[2]{\ensuremath{#1_0,\ldots,#1_{#2-1}}}

\newcommand{\id}{\ensuremath{\mbox{id}}}
\newcommand{\argt}{\ensuremath{\mbox{argt}}}
\newcommand{\prd}{\ensuremath{\mbox{prod}}}
\newcommand{\sq}{\ensuremath{\mbox{sq}}}
\newcommand{\tr}{\ensuremath{\triangleright}}

\newcommand{\para}{\vspace{0.05in}}
\newcommand{\shr}{\vspace{-0.1in}}
\newcommand{\lftrrw}{\mbox{ {\tt :-} }}

\section{Introduction}

This paper concerns the semantics of the part of Prolog
that remains when the built-in predicates have been removed
and when unification is enhanced by the occurs check.
Let us call this part ``pure Prolog''.
It can be regarded as the result
of Kowalski's procedural interpretation
of positive Horn clauses \cite{kwl79,kwl74,ckp72}.
The semantics of pure Prolog has been given by proof theory,
by model-theory, and by a fixpoint method \cite{vkw76,lld87,aptHandbook}.
All three approaches follow the syntax of clausal form.
As a result, the procedural interpretation has been ignored.
The purpose of the present paper is to remedy this defect.

One of the symptoms of the current deficiency in the semantics
of Prolog is that  procedures can only be recognized in an informal way.
As it stands, the procedural interpretation
does not provide procedure-valued expressions that can be
substituted for the procedure symbol in a procedure call.
Procedures are not ``first-class citizens''
the way functions can be in functional programming \cite{sto77}. 

\emph{Compositional semantics} does provide this possibility.
According to this method,
programs are expressions,
consisting, if composite, of an operation and its operand(s).
The value of the composite expression is the result of the operation
on the values of its operands.
The method is taken for granted when doing school-room sums:
the value of $(4 \div 2) \times (1+1)$ is $4$
because the value of $4 \div 2$ and $1+1$ are both $2$
and because $2 \times 2 = 4$.
In the late sixties Landin \cite{lnd63} and Scott and Strachey \cite{sst71}
applied the method to expressions that are programs.

In logic programming, compositional semantics seems to have been
used only for elucidating how the union of two logic programs
affects the definition of a predicate \cite{cddbgc93,bry96}.
In this paper we identify the compositions
that occur within a clause
and give a compositional semantics for these.

\para
There are several advantages to a compositional semantics
for a programming language.
One is that it guides implementation.
In fact, ``syntax-directed compilation'' \cite{rns61},
a widely used implementation technique,
is compositional semantics \emph{avant la lettre}.
The compositional semantics presented here decomposes logic programs
down to single procedure symbols, which take relations as value.
This accommodates relations that are not defined in the logic program
itself.

Another advantage of compositional semantics is that it forces
\emph{a language to be modular}.
For example, in a functional language with compositional semantics
$E_0E_1$ is the result of applying the value of $E_0$, which
must be a function, to the value of $E_1$, which may or may not be
a function.
The result can be a function, but need not be.

Compositionality  requires that the value of $E_0E_1$ does not change
when $E_0$ is replaced by a different expression with the same value.
This forces modularity in the sense that names of auxiliary functions
occurring in $E_0$ do not affect its value, hence are local.
Compositional semantics endows logic programs with the same property.
The value of a procedure call $p(t_0,\ldots,t_{n-1})$ is obtained
by an operation on the value of $p$ (which is a relation) and the
argument tuple $\langle t_0,\ldots,t_{n-1}\rangle$.
Again, the result depends on the value of the relational expression
substituted for $p$, not on the expression itself.

\paragraph{Contributions of this paper}

When one attempts a compositional semantics for the procedural
interpretation of logic, it becomes apparent that
it needs development beyond Kowalski's original formulation.
This is done in Section~\ref{procInt}.

Section~\ref{comSemLog} contains no contributions.
It needs to be included because cylindric set algebras are a compositional
semantics for first-order predicate logic and hence are a candidate for
compositional semantics for the procedural interpretation of logic.
This section includes enough to show why these algebras are not suitable.
We do find, however, an interesting connection between the tables
introduced here and the cylinders of Tarski (see Theorem~\ref{tabCyl}).

Tables, their operations and some of their properties are described in
Section~\ref{opTab}.
This is the basis on which the compositional semantics of
Section~\ref{compSem} rests.
Implications for modularity are discussed in Section~\ref{impMod}.

\section{Notation and terminology}
\label{mathObj}

In this section we collect terminology and notation that may differ
between authors.

\subsection{General terminology}
\begin{definition}[tuple, function, index set, type, restriction, subtuple]
--
A \emph{tuple} is a function $t$
that maps every index $i$ to $t(i)$,
which is called the tuple's component at $i$. \\
--
A \emph{function} is a triple
consisting of a set that is its \emph{domain},
a set that is its \emph{co-domain},
and a \emph{mapping} that associates
with every element of the domain
a unique element of the co-domain. \\
--
If the function is a tuple,
then the domain is usually called ``index set''. \\
--
The set of all functions with domain $S$ and co-domain $T$
is denoted $S \rightarrow T$. This set is often referred to as the
\emph{type} of the functions belonging to it. \\
--
Let $f$ be a function in $S\rightarrow T$ and let $S'$ be a subset of
$S$. $f \downarrow S'$ is the \emph{restriction} of $f$ to $S'$.
It has $S'$ as domain, $T$ as co-domain and its mapping associates
$f(x) \in T$ with every $x \in S'$. \\
--
If $t$ is a tuple with index set $I$ and if $I'$ is a subset of $I$,
then $t\downarrow I'$ is the \emph{subtuple} of $t$ defined by $I'$.
\end{definition}
\begin{definition}[relation]
A \emph{relation} with index set $I$ and co-domain $T$
is a set of tuples that have $I$ as index set
and $T$ as co-domain.
An $n$-ary relation is a relation that has
as index set the set
$\{0,\ldots,{n-1}\}$ of integers.
\end{definition}
Note that a relation need not be an $n$-ary relation.
Indeed, \emph{any set can be the index set of a relation}.

\begin{definition}[projection, cylindrification]
\label{projCyl}
Let $r$ be a relation that has $I$ as index set.
Let $I'$ be a subset of $I$.
The \emph{projection} $\pi_{I'}(r)$ of $r$ \emph{on}  $I'$
is $\{t \downarrow I' \mid t \in r\}$.

The \emph{cylinder in} $I$ \emph{on} a relation $r'$ with index set $I'$
is denoted $\pi_I^{-1}(r')$ and is
the greatest relation with index set $I$
and co-domain $T$ that has $r'$ as its projection
on $I'$; that is
$$ \pi_I^{-1}(r') =
  \cup \{\rho
  \mid \pi_{I'}(\rho) = r'
       \mbox{ and } \rho \mbox{ has index set } I
       \mbox{ and co-domain } T
       \}
$$
\end{definition}

\subsection{Mathematical objects arising in connection with
the semantics of logic programs}

To serve as semantic objects,
three basic objects are defined independently of another;
all three are mutually disjoint sets:
\begin{itemize}
\item
\Herb, an Herbrand universe
\item
\Var, a set of variables
\item
\Pred, a set of predicate symbols, also called ``procedure symbols''
\end{itemize}

From the three basic objects the following are derived:
\begin{itemize}
\item
$\T_V$, the set of terms that contain no function symbols or constants
other than those occurring in \Herb\ and no variables
other than those occurring in a subset $V$ of  \Var. 
We write \T\ for $\T_\Var$.
\item
Substitutions, each of which is a tuple
of type $V \rightarrow \T_V$,
for some subset $V$ of \Var.
If $\theta$ is a substitution and $\theta(x) = t$,
then we say that $\theta$ \emph{substitutes} $t$ \emph{for} $x$.
We may equate $\theta$ with the set
$\{x = t \mid \theta(x) = t \mbox{ and } x \in V\}$
of term equations.
\item
\emph{Term equations} are equations of the form $t_0 = t_1$,
where $t_0$ and $t_1$ are terms belonging to \T.
A set of term equations is said to be in \emph{solved form}
if every left-hand side is a variable, and if all these variables
are different, and if all variables in the right-hand sides also
occur as a left-hand side.
If a set of term equations has a solution, then it has a solution
in solved form. 

We will not distinguish between term equations in solved form,
substitutions, and tuples of elements of $T$ with a subset of
$\Var$ as index set.
\item
Relations consisting of tuples of elements of \Herb\ that are
indexed by $\{0,\ldots,n-1\}$. To distinguish these from the
next item, we refer to them as \emph{integer-indexed relations}.
\item
Relations consisting of tuples of elements of \Herb\ that are
indexed by a subset $V$ of \Var\
that is characteristic of the relation.
We refer to these as \emph{variable-indexed relations}.
\item
The  \emph{Herbrand base}, which is the set of ground atoms.
\item
\emph{Herbrand interpretations}, which are subsets of the Herbrand
base.
\item
\emph{Relational interpretations}, which are tuples of integer-indexed
relations indexed by \Pred.
\end{itemize}

\subsection{Compositional semantics}
Compositional semantics assigns the \emph{value}
$\M(E)$ to the \emph{expression} $E$. 
We are interested in expressions that are programs.
In this case the value is the behaviour of the program.
As ``value'' and ``behaviour'' do not match very well,
we often use ``meaning'' instead of ``value'' as a more neutral
term. It also happens to fit well with ``semantics''.

Compositionality of the semantics means that if $E$ is composed of
subexpressions $E_0$ and $E_1$, then $\M(E)$ is the result of
an operation on $\M(E_0)$ and $\M(E_1)$.
A well-known example illustrates
the compositional semantics of binary numerals.
It specifies how integers are assigned
as meanings to binary numerals:
$$\M(\verb+0+) = 0;
\M(\verb+1+) = 1;
\M(\verb+N0+) = 2 \M(\verb+N+);
\M(\verb+N1+) = 2 \M(\verb+N+) +1
$$

\section{The procedural interpretation of positive Horn clauses}
\label{procInt}

\subsection{The original procedural interpretation}

Kowalski \cite{kwl79} gives the procedural interpretation
of positive Horn clauses as follows:
\begin{quote}
``A Horn clause $B \leftarrow A_1,\ldots,A_m$, with $m \geq 0$, is interpreted as a
\emph{procedure} whose \emph{body} $\{A_1,\ldots,A_m\}$ is a set of procedure calls
$A_i$. Top-down derivations are \emph{computations}. Generation of a new goal
statement from an old one by matching the selected procedure call with the \emph{name}
$B$ of a procedure $B \leftarrow A_1,\ldots,A_m$ is a \emph{procedure invocation}.

A \emph{logic program} consists of a set of Horn clause procedures and is activated by
an initial goal statement.''
\end{quote}

Its semantics can be given by the fact that
a ground substitution $\theta$ is included in a result
of activating program $P$ with goal $G$
iff $P \cup \{G\theta\}$ is false in all Herbrand interpretations.
A more general characterization exists.

\subsection{A complete procedural interpretation}
The procedural interpretation of logic
can be formalized by expressing it as an \emph{abstract procedural syntax}.
Kowalski proposed, in effect,
$
B \leftarrow A_0,\ldots,A_{m-1}
$
as an alternative syntax
in the form of a decomposition of
$
\{B, \neg A_0,\ldots,\neg A_{m-1}\}
$
into a procedure heading and a procedure body.
This omits several decomposition steps:
(1)
the clause may be but one
of several several that can respond to the same procedure call,
so it is really a \emph{partial} procedure,
(2)
a body needs to be decomposed into calls, and
(3)
each call needs to be decomposed into its predicate symbol and its
argument tuple.
To make the procedural interpretation not only formal,
but also to complete it,
we propose Definition~\ref{procedural} as the abstract syntax
needed for compositional semantics.

\begin{definition}[procedural program]
\label{procedural}
\begin{center}
\begin{enumerate}
\item
A procedural program is a tuple of procedures
with index set \Pred\footnote{
The procedure symbols in \Pred\ index only one procedure.
This differs from Prolog where predicate symbols
include an arity.
}.
\item
A $n$-ary procedure is a set of $n$-ary clauses.
\item
An $n$-ary clause is a pair consisting of a parameter tuple of order $n$
and a procedure body.
\item
A procedure body is a set of procedure calls.
\item
A procedure call is a pair consisting of an $n$-ary procedure symbol
and an argument tuple of order $n$.
\item
A parameter tuple of order $n$ and
an argument tuple of order $n$ are both
$n$-tuples of terms.
\end{enumerate}
\end{center}
\end{definition}

Let us consider as example a set
$\Pred = $ \verb+{app,mem}+ of procedure symbols
and the procedural program in Figure~\ref{appMem};
let us call it $p$.
As $p$ is a tuple with \Pred\ as index set, and as a tuple
is a function, $p$ can be specified by\\
$p(\mbox{{\tt app}}) = $
\verb+{(nil,y,y) :- {}, (u.x,y,u.z) :- {app(x,y,z)}}+
\\
$p(\mbox{{\tt mem}}) = $ 
\verb+{(x,y) :- {app(u,x.v,y)}}+ \\

\begin{figure}
\begin{center}
\begin{tabular}{lcccl}
\hline
                                     &&&&                            \\
\verb+app(nil,y,y).+                 &&&& \verb+{app{(nil,y,y) :- {}+ \\
\verb+app(u.x,y,u.z) :- app(x,y,z).+ &&&& \verb+    ,(u.x,y,u.z) :- {app(x,y,z)}+ \\
                                     &&&& \verb+    }+ \\
\verb+mem(x,y) :- app(u,x.v,y).+     &&&& \verb+,mem{(x,y) :- {app(u,x.v,y)}}+\\
                                     &&&& \verb+}+                           \\
                                     &&&&                            \\
\hline
\end{tabular}
\end{center}
\caption{
\label{appMem}
A Prolog program (left) and an equivalent procedural program (right).
}
\end{figure}

By itself, Definition~\ref{procedural} defines \emph{some} procedural
language. It is only of interest in so far as it is related to
clausal logic. Similarly, the relational interpretations for procedural
programs need to be related to Herbrand interpretations.
Hence the following definition.

\begin{definition}[correspondence between logic and procedural programs]
\label{correspondence}
An Herbrand interpretation $I$ and a relational interpretation $R$
\emph{correspond} to each other ($I \sim R$) iff the following holds:\\
$
R(p) = \{\langle a_0,\dots,a_{n-1}\rangle \mid p(a_0,\dots,a_{n-1}) \in I  \}
$
for all $p \in \Pred$ and \\
$
I = \{p(a_0,\dots,a_{n-1})
     \mid p \in \Pred  \mbox{ and }\langle a_0,\dots,a_{n-1} \rangle  \in R(p)  \}
$

Let $S$ be a sentence consisting of positive Horn clauses (for which we assume
Kowalski's notation). Let $P$ be a procedural program. $S$ and $P$ \emph{correspond}
to each other ($S \sim P$) iff the following holds:\\
$
P(p) = \{ \mbox{partuple {\tt :-} body}
       \mid p(\mbox{partuple}) \leftarrow \mbox{ body} \in S\}
$
for all $p \in P$ and\\
$
S = \{ p(\mbox{partuple}) \leftarrow \mbox{ body}
    \mid
    \exists p \in \Pred \mbox{ such that } \mbox{partuple {\tt :-} body} \in P(p)
    \}
$.
\end{definition}

Each of the syntactical rules of Definition~\ref{procedural}
specifies that a certain type of expression is composed of
sub-expressions.
Compositional semantics then assigns to each of syntactical rules
a semantical rule that  specifies the corresponding operation
on meanings of the constituent sub-expressions.

The next section introduces the mathematical objects that are
suitable meanings.
Section~\ref{compSem} describes the semantical rules.

Before starting on this we give an informal idea of what is involved.
Let us work through the items in Definition~\ref{procedural},
starting at the bottom.
\paragraph{}

\begin{itemize}
\item[Rule 6]
Here we have very little to add:
a term denotes the set of its ground instances;
a tuple of terms denotes a tuple of sets of ground instances.
\item[Rule 5]
Consider the atoms
$p(x,v,w)$ and $p(u,w,y)$.
Although both involve the same relation $p$, they are different
calls and typically have different meanings.
These meanings are the result of a binary operation
with the relation $p$ and the tuple of arguments as operands.

The meaning of the entire call can be viewed as a selection
from the tuples that constitute relation $p$.
The selection is specified by the argument tuple,
and selects the tuples from the relation
that match the argument tuple.
Each such match takes the form of a substitution
for the variables in the argument tuple.
Therefore the result of the operation,
which we call \emph{filtering},
is a set of such substitutions.

As such sets are best presented in tabular form, we call
the result of the filtering operation on a relation
and an argument tuple  a \emph{table} (see Definition~\ref{tabDef}).
\item[Rule 4] 
We define the \emph{product} operation on tables
(see Definition~\ref{prodDef})
by means of which procedure bodies obtain values.
These values are tables.
Theorem~\ref{tabCyl} shows how product is related
to the semantic counterpart of conjunction
in Tarski's cylindric set algebra.
\item[Rule 3]
The meaning of a \emph{clause} is the $n$-ary relation
that results from an operation on the meanings
of the constituents of the clause:
the parameter tuple and the body.
As a parameter tuple has itself as meaning,
we define an operation,
which we call \emph{projection},
on a parameter tuple of order $n$ and a table
(see Definition~\ref{projDef}).
The operation yields an $n$-ary relation.
\end{itemize}

This completes the preview of the novel semantic operations:
filtering, product, and projection.
The remaining operations, those arising from Rules 1 and 2,
will not require any explanation beyond the following few lines.
In Rule~2, a procedure symbol is combined with a set of clauses.
As the meaning of a clause is an $n$-ary relation,
a set of such clauses denotes the union of these relations,
that is, an $n$-ary relation again.
Rule~2 merely creates a pair consisting of a procedure symbol
and a relation.

Rule~1 combines into a set a number of procedures,
each of which is a pair of a procedure symbol and a relation.
The semantic object corresponding to a program is therefore
a tuple of procedures indexed by \Pred, the set of procedure symbols.

\section{Compositional semantics for logic}
\label{comSemLog}

Though there does not seem to exist any compositional semantics
for the procedural interpretation of logic,
one does exist for logic that is parsed in the conventional way.
It is called \emph{algebraic logic},
which would be called compositional semantics
if it would concern a programming language.
It is therefore a good starting point
for a compositional semantics of logic programs.

Algebraic logic assigns elements of an algebra
as meanings to formulas of logic;
it assigns operations of the algebra
as meaning to the connectives that compose logical formulas.
The more widely known approach to algebraic is based on
the cylindric set algebras of Tarski \cite{hmt85,trsk52}
of which we give a brief sketch here.
Tarski's approach is based on the algebraic interpretation
of propositional logic due to Boole \cite{bl54}.

\subsection{Propositional logic and Boolean algebra}
In general, a Boolean algebra is any algebra
that satisfies certain defining axioms. A Boolean \emph{set} algebra is a special
case. It is described as the tuple
$
\langle S,\cup,\cap,\sim,\emptyset,U \rangle
$
where $S$ is a set of subsets of $U$
that contains $\emptyset$ and $U$
and is closed under union, intersection,
and complementation (here denoted as $\sim$).

A special case of a Boolean set algebra
is the one where $U$ is the Cartesian product $D^n$,
for some given non-empty set $D$.
Recall that the Cartesian product $D^n$
is the set of all $n$-tuples of elements of $D$.
We can further specify the Boolean set algebra
by choosing $U = D^0 = \{\langle \rangle\}$
and $S= \{\{\}, \{\langle \rangle\}\}$.
As a result, the algebra has two elements: 
$\{\}$ and
$\{\langle \rangle\}$.
Boolean addition, multiplication, and complementation
then become set union, set intersection, and set complement, respectively.
Let \M\ be the mapping from propositional formulas
to the elements of the Boolean algebra.
We have that
$\M(p_0 \vee p_1) = 
\M(p_0) \cup  \M(p_1)
$,
$\M(p_0 \wedge p_1) = 
\M(p_0) \cap \M(p_1)
$, and
$\M(\neg p) = \;\sim \M(p) $
when  we define 
$\M(true) = \{\langle \rangle\}$
and
$\M(false) = \{\}$.

\subsection{Predicate logic and cylindric set algebra}
Tarski sought an algebra that would do for first-order predicate logic
what Boolean algebra does for propositional logic.
The  result was cylindric set algebra \cite{trsk52,hmt85}.

In model theory, formulas correspond to relations.
If this intuitively attractive feature is to be retained,
a puzzle needs to be solved.
Consider $\M(p(x,y) \wedge p(y,z))$.
As the formula has three free variables, this should be 
a ternary relation.
As conjunction means the same in predicate logic as in propositional logic,
this ternary relation should be the result of set intersection.
But the arguments of the set intersection
are derived from binary predicates.

Another part of the puzzle is that
$p(x,y)$ and $p(y,z)$ should both denote binary relations,
but these should be different
and cannot both be the relation denoted by $p$.

Tarski solved these conundrums by mapping every formula
to a relation consisting tuples indexed by \emph{all} the variables
in the language.
He assumed a countable infinity of variables in the language,
in a given order.
In this way he could identify each variable with a natural number.
Thus this meaning algebra has as elements
relations that are subsets of the Cartesian product $D^\omega$.

The choice of the two $0$-ary relations on $D$ for the two elements
of the Boolean algebra for propositional logic is now clear:
the number of variables in a propositional formula is $0$.

A first-order predicate logic formula without free variables
is either true or false.
It is mapped accordingly to the full or empty $\omega$-ary relation over $D$;
that is, to $D^\omega$ or $\emptyset$.
At first sight it might seem right to map a formula
$F[x_0,\ldots,x_{n-1}]$
with free variables $x_0,\ldots,x_{n-1}$
to the relation that consists of all the
tuples
$\langle a_0,\ldots,a_{n-1}\rangle$
such that
$F[a_0,\ldots,a_{n-1}]$
is true.
By mapping instead this formula
to the \emph{cylinder} on this relation with respect to all variables,
Tarski ensured that
$\M(p_0 \vee p_1) =
\M(p_0) \cup \M(p_1)
$
and
$\M(p_0 \wedge p_1) =
\M(p_0) \cap \M(p_1)
$,
just as in the case of propositional logic.

Going back to the above puzzle,
we see that
$\M(p(x,y))$
and
$\M(p(y,z))$
are not binary relations
but $\omega$-ary relations that are
cylinders on a binary relation.
Though the binary relation denoted by $p$ in
these formulas is the same,
the cylinders on $\M(p(x,y))$ and $\M(p(y,z))$ are different.
In this way $\M(p(x,y)) \cap \M(p(y,z))$
is a cylinder on a ternary relation.

Thus Tarski devised a compositional semantics for first-order predicate logic.
He simplified the language to contain as connectives
only conjunction, disjunction, and negation.
The presence of the negation connective
makes it possible to do with a single quantifier,
the existential one.
There are no function symbols.
An atomic formula can be of the form $x = y$.

For this language a suitable algebra for a compositional semantics is
the \emph{cylindric set algebra}
$
\langle S, \cup, \cap, \sim, \emptyset, D^\omega, C_k, \delta_{i,j}\rangle
$
for all natural numbers $i$, $j$, and $k$.
This algebra is a Boolean algebra (for the first six items).
In addition, there are
$\delta_{i,j}$, the $(i,j)$ diagonal relations:
the subsets of $D^\omega$ consisting of the tuples
where the elements indexed by $i$ and $j$ are equal.
The specification of cylindric set algebras also includes
for all $k \in \omega$
the cylindrification operations $C_k$, which are defined
by $C_k r$ being the subset of $D^\omega$
consisting of the tuples
that differ from a tuple in $r$ in at most the $k$-th component.

$S$ is the set that contains $\emptyset$, $D^\omega$,
as well as all the diagonal relations $\delta_{i,j}$
and that is closed under the Boolean operations
as well as under $C_k$.

\subsection{Cylindric set algebra
            for the compositional semantics of procedural programs?}
Cylindric set algebra interprets formulas as relations;
relations are a suitable model for the procedures
of a procedure-oriented language.
These facts might suggest that cylindric set algebras be used for
a compositional semantics for the procedural interpretation of logic.

The following are reasons not to do so.
\begin{itemize}
\item
Tarski's choice of language for first-order predicate logic
is no more procedure-oriented than clausal form is.
\item
Tarski's semantics does not specify by what operation,
for example, the binary relation $\M(p(x,y,x))$
arises from the ternary relation $p$
and the argument tuple $\langle x,y,x \rangle$.
That is, his compositionality stops short of the atomic formula.
\end{itemize}
Accordingly, we create an independent alternative,
centered around the concept of \emph{table}.
Surprisingly, one of the operations on tables
reflects the way Tarski uses cylinders
to algebraize the logical connectives.

\section{Tables}
\label{opTab}

Some of the semantic objects
for the procedural programs of Definition~\ref{procedural} 
are familiar; they have been
introduced in Section~\ref{mathObj}.
This section is devoted to the one novel type of semantic object.

\begin{definition}[table]
\label{tabDef}
A table on a subset $V$ of \Var\ is a set of tuples
each of which has type $V \rightarrow \T_V$.
If the set of tuples is empty, then we have the \emph{null table},
which we write as $\bot$.
If $V$ is empty and the set of tuples is not, then the table
is the \emph{unit table}, which we write as 
$\top$.
\end{definition}
As there is only one function of type $\{\} \rightarrow \T_V$
for any subset $V$ of \Var,
we have that $\top = \{\langle \rangle\}$.

To every table there corresponds
a unique variable-indexed relation,
which we call the result of \emph{grounding} the table.
\begin{definition}[grounding, table equivalence]
Let $t$ be a table with tuples of type $V \rightarrow \T_V$.
$\Gamma(t)$, the result of \emph{grounding} $t$,
is the variable-indexed relation
consisting of the tuples of type $V \rightarrow \Herb$
each of which is a ground instance of a tuple in $t$.

Tables $t_0$ and $t_1$ are \emph{equivalent}
if $\Gamma(t_0) = \Gamma(t_1)$.
\end{definition}

In this section we define and discuss
the product, filtering, and projection operations.
These operations are adapted
from \cite{vbr91}, where filtering is called ``application''.

\subsection{Product}
\label{prodSec}

As we will see,
compositional semantics assigns tables as values
to the calls in a procedure body
as well as to the body itself.
The co-occurrence of calls in a body
corresponds to the \emph{product} operation
of the corresponding tables.
An example will be given in Section~\ref{comprExample}.

\begin{definition}[product]
\label{prodDef}
Let $\tau_0$ and $\tau_1$ be tables
consisting of tuples with index
sets $V_0$ and $V_1$, respectively.
The \emph{product} $\tau_0 \ast \tau_1$ of these tables is defined
as a table with $V_0 \cup V_1$ as
index set. 
The product table $\tau_0 \ast \tau_1$
contains a tuple $t$ if and only if
there is a tuple $t_0$ in $\tau_0$
and
a tuple $t_1$ in $\tau_1$
such that the set of equations $t_0 \cup t_1$
is solvable and has $t$ as solved form.
\end{definition}

\begin{theorem}
--
Product is commutative and associative. \\
--
The null table $\bot$ is an absorbing element: 
$\bot \ast \tau = \tau \ast \bot = \bot$ for all tables $\tau$. \\
--
The top table $\top$ is a unit:
$\top \ast \tau = \tau \ast \top = \tau$ for all tables $t$. \\
--
$\tau\ast \tau$ and $\tau$ are equivalent.
\end{theorem}

Commutativity and associativity give the obvious
meaning to $\ast S$, where $S$ is a set of tables,
assuming that $*\{\} = \top$.

\begin{definition}[cylinder on table]
\label{cylTab}
The cylinder $\pi^{-1}(T)$ on a table $T$ with index set $V \in \Var$
is a table where \Var\ is the index set
and where every tuple $t'$ is obtained
from a tuple $t$ in $T$ by defining
$t'(v) = t(v)$ for every $v \in V$ and
$t'(v) = v$ for every $v \in \Var \setminus V$.
\end{definition}
This definition of ``cylinder'' is independent of Tarski's notion,
which is the one in Definition~\ref{projCyl}.
The two notions are connected as follows.
\begin{lemma}
Let $T$ be a table with index set $V$, a subset of \Var.
We have that
$
\Gamma(\pi^{-1}(T)) =
\pi^{-1}(\Gamma(T))
$.
The first occurrence of $\pi^{-1}$ is the cylindrification
on tables from Definition~\ref{cylTab};
the second one is the cylindrification on relations
in Definition~\ref{projCyl}.
\end{lemma}
The distinguishing feature of Tarski's use of cylindric set algebra
as semantics for first-order predicate logic is that conjunction
in logic simply translates to intersection in the algebra.
And this is the case even though the conjunction may be between
two formulas with sets $V_0$ and $V_1$ of free variables.
There is no restriction on these sets: they may be disjoint,
one may be a subset of the other, or neither may be the case.
Tarski's device works
because the intersection is not between
relations with $V_0$ and $V_1$ as index sets,
but between \emph{cylinders} on these relations
in the set of all variables.
This crucial idea reappears in the product of tables defined here.
The connection is made apparent by the following theorem.
\begin{theorem}
\label{tabCyl}
Let $\tau_i$ be a table with set $V_i$ of variables, for $i \in \{0,1\}$.
$\Gamma(\tau_0 \ast \tau_1)
= \pi_{V_0 \cup V_1}
(\pi^{-1}_{\Var}(\Gamma(\tau_0)) \cap \pi^{-1}_{\Var}(\Gamma(\tau_1)))
$.
\end{theorem}

\subsection{Filtering: from relations to tables}
\label{filtSec}
Just as in a functional programming language a function is applied
to the $n$-tuple of its arguments,
we think of the combination of a procedure symbol with its argument
tuple as a binary operation.
Consider therefore a call consisting of a procedure symbol
and an argument tuple of order $n$.
The procedure symbol has as value an integer-indexed relation
of order $n$.
It combines with the argument tuple to produce a table.
This is the operation we call \emph{filtering}.
An example of this operation can be found in Section~\ref{comprExample}.
\begin{definition}[filtering]
Let $p$ be an integer-indexed relation of order $n$
and let $t$ be an $n$-tuple of terms with $V$ as set of variables.
The result of the \emph{filtering} $p:t$ is a table
where $V$ is the index set of the tuples.
For every tuple $\langle a_0, \ldots, a_{n-1} \rangle$
in $p$ for which the set
$\{t_0 = a_0, \ldots, t_{n-1} = a_{n-1}\}$
of equations is solvable,
the table contains a tuple that is the solved form of these equations.
\end{definition}
In functional programming, an expression $E_0E_1$ denotes
function application. Here $E_0$ is an expression that evaluates
to a function, and it is this function that is applied.
Filtering is the relational counterpart:
in $p:t$ the first operand $p$ has a relation as value;
it is filtered by the tuple $t$;
the result is a table.

\subsection{Projection: from tables to integer-indexed relations}
\label{projSec}

Finally, a clause is a contribution to a procedure,
which is an integer-indexed relation of order $n$.
This relation, which is the clause's value,
is somehow produced by a combination of
the parameter tuple of the clause and the table that is the value of its body.
We call this operation \emph{projection}.
An example of this operation can be found in Section~\ref{comprExample}.

\begin{definition}[projection]
\label{projDef}
Let $T$ be a table consisting of tuples whose index set
is a subset $V$ of \Var\ .
The result of \emph{projecting} $T$ \emph{on} an $n$-tuple of terms,
denoted 
$\langle t_0,\ldots,t_{n-1} \rangle / T$,
is an integer-indexed relation consisting of $n$-tuples of ground terms.
The relation contains such a tuple if and only if it is
a ground instance of
$\langle t_0\theta,\ldots,t_{n-1}\theta \rangle$,
for some $\theta$ in $T$.
\end{definition}

Usually every variable in the parameter tuple of a clause
occurs also in the body of that clause.
It often happens
that the parameter tuple contains
fewer variables than the body.
For example, \verb+mem(x,y) :- app(u,x.v,y)+.
Any operation that yields an integer-indexed relation
consisting of $n$-tuples from a body
that contains more than $n$ variables
is reminiscent of a projection operation.
Hence the name.

\subsection{Are projection and filtering inverses?}

Now that we have operations from tables
to relations and vice versa, one
may wonder whether these are each other's inverses.
The short answer is, in general, ``no'', because
$$((t_0,\ldots ,t_{n-1})/T) : (t_0,\ldots ,t_{n-1})$$
is not always the table $T$. Take, for example, the case that
$t_0,\ldots ,t_{n-1}$ have no variables.
Then the above expression is $\top$ whenever $T$ is not $\bot$.
But the absence of variables in $t_0,\ldots,t_{n-1}$ is a
rather pathological case. When we add certain restrictions, we can say
that, in a sense, ``/'' and ``:''
are each other's inverses, as shown by the following theorems.

\begin{theorem}
\label{exact}
For all tables $T$ with a subset $V$ of \Var\ as index set
and all terms $t_0,\ldots ,t_{n-1}$ with $V$ as set of variables
$((t_0,\ldots ,t_{n-1})/T) : (t_0,\ldots ,t_{n-1})$
is equivalent to $T$.
\end{theorem}

For an inverse in the other direction,
compare the $n$-ary relation $r$ with
$$(t_0,\ldots ,t_{n-1})/(r : (t_0,\ldots ,t_{n-1})).$$
That this expression does not always equal $r$ is shown by
$(c,d)/(\{(a,b)\} : (c,d)) = \{\},$
where $a$, $b$, $c$ and $d$ are constants.
This example suggests:

\begin{theorem}
For all $n$-ary relations $r$ and all terms $t_0,\ldots ,t_{n-1}$, we have
$$(t_0,\ldots ,t_{n-1})/(r : (t_0,\ldots ,t_{n-1})) \subseteq r.$$
\end{theorem}

However, by strengthening the restrictions, we can have equality instead of
inclusion, as shown in the following theorem.

\begin{theorem}
For all $n$-ary relations $r$ and all distinct variables $x_0,\ldots,x_{n-1}$, we
have $$(x_0,\ldots ,x_{n-1})/(r : (x_0,\ldots ,x_{n-1})) = r.$$
\end{theorem}

\section{Compositional semantics}
\label{compSem}

The operations of product, filtering, and projection are intended to
be the semantical counterparts of the way in which procedural programs
are put together syntactically.
But so far only the intention exists.

The definition below formalizes this intention.
It defines the meaning $\M(P)$ of a procedural program $P$,
where $P$ is regarded
as a tuple with index set \Pred\ of integer-indexed relations.
This meaning depends
on a relational interpretation $I$ (Definition~\ref{correspondence})
that assigns relations to the procedure symbols in \Pred.
We indicate this dependence by a subscript, as in \MI.

Definition~\ref{MI} gives the compositional semantics for procedural
programs.
As Definition~\ref{correspondence} shows, a procedural program is
just another way of writing a set of positive Horn clauses.
The semantics of these has been defined in three equivalent ways:
model-theoretically, proof-theoretically, and by means of fixpoints.
The main theorem (\ref{main}) of this paper
relates the compositional semantics of procedural programs
to the established semantics of the corresponding clausal sentences.

\begin{definition}
\label{MI}
\begin{enumerate}
\item
For every procedural program $prog$,
$\MI(prog)$ is the  tuple with index set \Pred\
such that for every $prsym \in \Pred$
the $prsym$-component is  $\MI(prog(prsym))$.
\item
For every procedure $proc$,
$\MI(proc) = \cup \{\MI(clause) \mid clause \in proc\}$
\item
For every clause with $pars$ as parameter tuple and
$B$ as body,\\
$\MI(pars \mbox{ {\tt :-} } B) = pars / \MI(B)$
(use of projection)
\item
For every procedure body $B$, we have 
$\MI(B) = \ast \{\MI(call) \mid call \in B\}$
(use of product)
\item
For every call with prsym as procedure symbol and
args as argument tuple,
$\MI(prsym \; args) = \MI(prsym) :  args $
(use of filtering)
\item
For every $prsym \in \Pred$ we have that
$\MI(prsym) = I(prsym)$
\end{enumerate}
\end{definition}

Here the numbering follows that of the syntactical rules
of Definition~\ref{procedural}.
\begin{theorem}
\label{main}
Let $I$ be a relational interpretation and
$I'$ the corresponding (Definition~\ref{correspondence})
Herbrand interpretation.
Let $P$ be a procedural program and
$P'$ the corresponding (Definition~\ref{correspondence})
set of positive Horn clauses.
We have 
$$
T_{P'}(I') \sim \MI(P),
$$
where $T$ is the immediate-consequence operator for logic programs.
\end{theorem}

We only know a cumbersome,
though straightforward,
proof of this theorem.

$T$ has a unique least fixpoint \cite{vkw76,lld87,aptHandbook}.
The partial order among Herbrand interpretations (set inclusion)
translates according to the correspondence in Definition~\ref{correspondence}
to a partial order among relational interpretations (component-wise inclusion).
Hence there is, for each procedural program $P$,
a unique least relational interpretation $I$ such that
$I = \M_I(P)$.

\begin{definition}
$\M(P) = \M_{I_m}(P)$ where $I_m$ is the least relational interpretation $I$
such that  $I = \M_I(P)$.
\end{definition}

\begin{theorem}
Let $P'$ be a logic program and let $P$ be the corresponding procedural
program. Then we have $\mbox{lfp}(T_{P'}) \sim  \M(P)$.
\end{theorem}
This relates the compositional semantics of procedural programs to the mutually
equivalent least fixpoint, proof-theoretical, and model-theoretical semantics
of logic programs.

\subsection{An example}
\label{comprExample}

Consider the procedural program clause
$(f(y),z) \lftrrw \{p(x,f(y)), p(f(x),z)\}.$
Here $\M(p)$ is an integer-indexed relation with $\{0,1\}$ as
index set.
Let us assume that $\M(p) = $
\begin{tabular}{c||c|c|c|c|}
0 &  \va    & \vfa & \vfa & \vfb   \\
\hline 
1 &  \vfb & \vb  & \vfb & \vfa   \\
\end{tabular} .
Here the four 2-tuples, indexed by $\{0,1\}$,
are displayed vertically.

The value of a call is a table; that is, a variable-indexed relation.

$\M(p(x,f(y))) = \M(p) : \langle x,f(y) \rangle =$
\begin{tabular}{c||c|c|c|}
$x$ & \va & \vfa & \vfb   \\
\hline
$y$ & \vb & \vb    & \va     \\ 
\end{tabular} .

Similarly,
$\M(p(f(x),z)) = \M(p) : \langle f(x),z \rangle =$
\begin{tabular}{c||c|c|c|}
$x$ & \va & \va    & \vb       \\
\hline
$z$ & \vb & \vfb & \vfa    \\ 
\end{tabular} .

The value of the body is the product of the above two tables:\\
$\M(p(x,f(y)),p(f(x),z)) = \M(p(x,f(y))) \ast \M(p(f(x),z))= $
\begin{tabular}{c||c|c|}
$x$ & \va & \va    \\
\hline
$y$ & \vb & \vb    \\ 
\hline
$z$ & b & \vfb \\ 
\end{tabular} .

Finally, the meaning of the entire clause\\ 
$(f(y),z) \lftrrw {p(x,f(y)), p(f(x),z)}$
is obtained by projection: \\
$\langle f(y),z \rangle / \M(p(x,f(y)),p(f(x),z))=$ 
\begin{tabular}{c||c|c|}
$0$ &  \vfb & \vfb   \\
\hline 
$1$ &  \vb    & \vfb   \\
\end{tabular} .

\section{Implications for modularity}
\label{impMod}

Suppose $P$ and $P'$ are procedural programs
with the same Herbrand universe.
If $p(t_0,\ldots,t_{n-1})$ is a call in $P$,
then the meaning of $p$ is $(\M(P))(p)$.
But $p$ is a special case of an expression
that has an $n$-ary integer-indexed relation as value.
Such an expression could also be $(\M(P'))(p')$
if $p'$ is a procedure symbol in $P'$ paired with an $n$-ary procedure.
The value of this expression is a set of $n$-tuples of ground terms.
This value is independent of the procedure symbols occurring in $P'$.
Hence these symbols are ``encapsulated'' in the expression $(\M(P'))(p')$.

This only addresses the semantics of a module mechanisms.
It leaves open the syntax that indicates which set of clauses
is a module and which procedure symbol is exported. 

\section{Related work}

Modules for logic programs can be obtained
via proof theory \cite{mllr89,mccb92}
or via higher-order logic \cite{chkfwr93}.
A different approach is to base it
on decompositions of the immediate-consequence operator
as done by Brogi et al. \cite{brmnpdtr94}.
It is baffling that the various approaches to modularity  are so difficult
to relate.
Several more are mentioned by Brogi et al. \cite{brmnpdtr94},
who also seem at a loss in relating them to their own work.

Additional details about the operations on tables and relations,
there called ``table-relation algebra'',
can be found in \cite{vbr91,brhm92}.

\section{Concluding remarks}

The procedural programs of Definition~\ref{procedural}
are the result of the desire to give a procedural
interpretation not only of an entire clause, but also
of the composition of head and body within a clause
as well as of the compositions that can be recognized in the body.
Thus procedural programs are but another way of parsing
a set of positive Horn clauses.

But suppose that in 1972 one had never heard of clausal logic and that
the motivation was to characterize in what way languages
with procedures, such as Algol, are of a higher level than
their predecessors.
A higher level of programming in such languages is achieved by
using procedure calls as much as possible.
That suggests the ultimate altitude in level of programming:
procedure bodies contain procedure calls only.

What about data structures for a pure procedural language?
Just as Lisp simplified by standardizing all data structures
to lists, one could make a similar choice by standardizing on
trees.
In this way a pure procedure-oriented language would arise
that coincides with the procedural programs of Definition~\ref{procedural}.

Functional programming languages have an obvious semantics
in the form of functions as defined in mathematics.
The semantics of Algol-like languages is defined in terms of
transitions between computational states.
These transitions are specified directly or indirectly in terms
of assignments.
In this way one might think that procedure-oriented programming languages
are of inherently lower level than functional programming languages.

It is not necessary to specify procedures in terms of state transitions.
A procedure is more directly specified as the set of all possible
combinations of values of the arguments of a call.
That is, as a set of tuples of the same arity, which is a relation.

In this way the procedural programs of Definition~\ref{procedural} become
as high-level as functional programs and obtain a semantics that is as
mathematical.

One might argue that this gives procedural programs
a significance that extends beyond logic programming.
For example, they may be a way to describe Colmerauer's view \cite{clmr79}
that Prolog is not necessarily a logic programming language.
In the procedural interpretation described here,
the Herbrand universe can be replaced
by a sufficiently similar data structure,
such as the rational trees.

\section{Acknowledgements}
I am grateful to Belaid Moa and three anonymous reviewers
for their suggestions for improvement.
This research was supported by the University of Victoria and by
the Natural Science and Engineering Research Council of Canada.


\end{document}